\documentclass[12pt,a4paper]{article}
\usepackage[top=25truemm,bottom=35truemm,left=25truemm,right=25truemm]{geometry}
\usepackage{amsmath,amssymb}
\usepackage{graphicx}

\begin{document}
\setlength{\baselineskip}{18pt}
\begin{titlepage}
\begin{flushright}
\begin{tabular}{l}
 SU-HET-03-2015\\
\end{tabular} 
\end{flushright}

\vspace*{1.2cm}
\begin{center}
{\Large\bf Vacuum stability in the $U(1)_\chi$ extended model
 with vanishing scalar potential at the Planck scale 
}
\end{center}
\lineskip .75em
\vskip 1.5cm

\begin{center}
{\large Naoyuki Haba}$^1$ and
{\large Yuya Yamaguchi}$^{1,2}$\\

\vspace{1cm}

$^1${\it Graduate School of Science and Engineering, Shimane University,\\
 Matsue 690-8504, Japan}\\
$^2${\it Department of Physics, Faculty of Science, Hokkaido University,\\
 Sapporo 060-0810, Japan}\\

\vspace{10mm}
{\bf Abstract}\\[5mm]
{\parbox{13cm}{\hspace{5mm}

We investigate the vacuum stability in a scale invariant local $U(1)_\chi$ model
 with vanishing scalar potential at the Planck scale.
We find that
 it is impossible to realize the Higgs mass of 125\,GeV
 while keeping the Higgs quartic coupling $\lambda_H$ positive in all energy scales,
 that is, the same as the standard model.
Once one allows $\lambda_H<0$, 
 the lower bounds of the $Z'$ boson mass are obtained
 through the positive definiteness of the scalar mass squared eigenvalues,
 while the bounds are smaller than the LHC bounds.
On the other hand,
 the upper bounds strongly depend on
 the number of relevant Majorana Yukawa couplings of the right-handed neutrinos $N_\nu$.
Considering decoupling effects of the $Z'$ boson and the right-handed neutrinos,
 the condition of the singlet scalar quartic coupling $\lambda_\phi>0$
 gives the upper bound in the $N_\nu=1$ case,
 while it does not constrain the $N_\nu=2$ and 3 cases.
In particular, we find that
 the $Z'$ boson mass is tightly restricted for the $N_\nu=1$ case
 as $M_{Z'} \lesssim 3.7\,{\rm TeV}$. 
}}
\end{center}
\end{titlepage}

\section{Introduction}
The standard model- (SM-)like Higgs boson was discovered at the LHC,
 and its mass was obtained by the ATLAS and CMS combined experiments as
\begin{eqnarray}
	M_h =125.09 \pm 0.21 \ ({\rm stat.}) \pm 0.11 \ ({\rm syst.})\ {\rm GeV},
\label{Higgs}
\end{eqnarray}
 with a relative uncertainty of 0.2$\%$ \cite{Aad:2015zhl}.
The SM predicts that
 the quartic coupling of the Higgs $\lambda_H$ and its $\beta$ function $\beta_{\lambda_H}$
 becomes zero below, but close to, the Planck scale
 ($M_{\rm Pl} = 2.435 \times 10^{18}$\,GeV) \cite{Buttazzo:2013uya}.
The negative quartic coupling causes a vacuum stability problem,
 which may suggest the appearance of new physics below the Planck scale.
In fact, the vacuum of the Higgs potential is meta-stable in the SM,
 and the vacuum stability has been discussed in a number of works
 \cite{Holthausen:2011aa}--\cite{Haba:2014oxa}.
In particular, the multiple point principle (MPP) requires
 the vanishing $\lambda_H$ and $\beta_{\lambda_H}$ at a high energy scale,
 and it suggests a $135 \pm 9$\,GeV Higgs mass with the top pole mass as $173 \pm 5$\,GeV
 \cite{Froggatt:1995rt}
 (see also Refs. \cite{Froggatt:2001pa}--\cite{Kawana:2015tka} for more recent analyses).
Note that the conditions of the MPP could be naturally realized by the asymptotic safety of gravity
 \cite{Shaposhnikov:2009pv}.

The vanishing the Higgs quartic coupling near the Planck scale might suggest that
 the Higgs potential is completely flat at the Planck scale,
 and this possibility has been studied
 in Refs. \cite{Iso:2009ss}--\cite{Chun:2013soa}.
In this context,
 the Higgs mass term is forbidden by a classical conformal invariance.
The classical conformal invariance could be broken in general
 by radiative corrections via the Coleman-Weinberg (CW) mechanism \cite{Coleman:1973jx},
 or a condensation in a strongly coupled sector like the QCD.
In particular, in a flatland scenario, which is called in Ref.~\cite{Hashimoto:2013hta},
 an additional local $U(1)$ symmetry exists,
 and it is radiatively broken by the CW mechanism.
Then, since the SM singlet scalar gets a nonzero vacuum expectation value (VEV),
 its mixing term with the Higgs becomes the Higgs mass term.
If the mass term is negative,
 electroweak (EW) symmetry breaking could successfully occur.
In Ref.~\cite{Hashimoto:2014ela},
 the authors investigated the possibilities of the flatland scenario
 in various $U(1)$ extended models.

In addition,
 the hierarchy problem for the Higgs mass can be solved in the flatland scenario as follows.
From Bardeen's argument \cite{Bardeen:1995kv},
 the quadratic divergence of the Higgs mass can always be multiplicatively subtracted
 at some energy scale.
Once the mass term is renormalized at a high energy scale, e.g., the Planck scale,
 the quadratic divergence does not appear at lower energy scales.
Then, the hierarchy problem is an issue only for logarithmic divergences.
Since the renormalization group equation (RGE) of the Higgs mass term in the SM
 is proportional to itself,
 if it is zero at a high energy scale,
 it continues to be zero at lower energy scales as long as the theory is valid.
However, if there is a mixing term between the Higgs and other scalar field,
 the RGE of the Higgs mass term includes a term proportional to the scalar mass squared.
This term comes from the logarithmic divergence due to the loop diagram of the scalar field.
Then, the correction would be relevant for a realization of the Higgs mass
 when the scalar mass is not so large compared to the EW scale.
Therefore, the hierarchy problem can be solved
 if no large intermediate scales exist between the EW and the Planck scales.

In this paper,
 we begin with a review of the flatland scenario in Sect.~\ref{sec:model},
 in which we use the $U(1)_\chi$ extended model as in Ref.~\cite{Chun:2013soa}.
It is known that
 the CW mechanism can occur and the EW symmetry is successfully broken in this model
 (see Ref.~\cite{Hashimoto:2014ela}).
However,
 a running of the singlet scalar quartic coupling is quite different from the typically expected one,
 when the number of relevant Majorana Yukawa couplings of the right-handed neutrinos is two,
 i.e., $N_\nu=2$.
Nevertheless, we find that the CW mechanism can also successfully occur in the $N_\nu=2$ case.
Next, we investigate the vacuum stability using two-loop RGEs in Sect.~\ref{sec:vacuum}.
We find that
 it is impossible to realize the Higgs mass of 125\,GeV
 while keeping $\lambda_H>0$ at all energy scales,
 that is, the same as the SM.
Once one allows $\lambda_H<0$, 
 the lower bounds of the $Z'$ boson mass are obtained
 through the positive definiteness of the scalar mass squared eigenvalues,
 while the bounds are smaller than the LHC bounds.
On the other hand,
 the upper bounds strongly depend on $N_\nu$.
Considering the decoupling effects of the $Z'$ boson and the right-handed neutrinos,
 the condition of the singlet scalar quartic coupling $\lambda_\phi>0$
 gives the upper bound in the $N_\nu=1$ case,
 while it does not constrain the $N_\nu=2$ and 3 cases.
Finally, we mention the experimental bounds on the $Z'$ boson mass in Sect.~\ref{sec:ex-bound},
 and find that
 the $Z'$ boson mass is tightly restricted for the $N_\nu=1$ case
 to $2.24\ (2.59)\,{\rm TeV} \lesssim M_{Z'} \lesssim 3.7\,{\rm TeV}$,
 where the lower bound corresponds to the ATLAS (CMS) result.

\section{$U(1)_\chi$ extension of the SM in the flatland scenario} \label{sec:model}

We consider the $U(1)_\chi$ extension of the SM,
 in which field contents are as shown in Table \ref{particles}.
A scalar potential is given by
\begin{eqnarray}
	V = \lambda_H |H|^4 + \lambda_\Phi |\Phi|^4 + \lambda_{\rm mix} |H|^2 |\Phi|^2,
\end{eqnarray}
 where $H$ and $\Phi$ are a Higgs doublet and an SM singlet complex scalar, respectively.
Since we assume the classical conformality,
 there are no dimensional parameters such as mass squared terms.
In the flatland scenario,
 we impose that all the quartic couplings vanish at the Planck scale.
The Lagrangian including right-handed neutrinos $N$ is given by
\begin{eqnarray}
	{\cal L}_M = -Y_N^{\alpha i} \overline{L_\alpha} H N_i - Y_M^{ij} \Phi \overline{N_i^c}N_j + (h.c.),
\end{eqnarray}
 where $L$ is the lepton doublet,
 and $\alpha$ and $i$ show the indices of the flavor and mass eigenstates, respectively.
Since the type-I seesaw mechanism generates the active neutrino masses
 by integrating out right-handed neutrinos with TeV-scale masses,
 the Dirac Yukawa couplings are typically ${\cal O}(10^{-6})$.
Thus, we neglect $Y_N$ for the RGE analyses in the following.
Here, there are two $U(1)$ gauge bosons,
 and we take their kinetic terms as diagonal.
Then, the covariant derivative is given by
\begin{eqnarray}
	D_\mu = \partial_\mu - ig_3 T^\alpha G^\alpha_\mu - ig_2 \tau^a W^a_\mu
		- ig_Y YB^Y_\mu - i(g_{\rm mix} Y + g_\chi X)B^X_\mu,
\end{eqnarray}
 where $Y$ and $X$ denote $U(1)_Y$ and $U(1)_\chi$ charges, respectively.
The $U(1)_\chi$ gauge boson is conventionally called the $Z'$ boson,
 and we denote $Z'_\mu \equiv B^X_\mu$ hereafter
 (see Ref.~\cite{Langacker:2008yv} for a review of the $Z'$ boson).
The gauge couplings of $SU(3)_c$, $SU(2)_L$, $U(1)_Y$, and $U(1)_\chi$
 are $g_3$, $g_2$, $g_Y$, and $g_\chi$, respectively.
In addition, there is a $U(1)$ mixing coupling $g_{\rm mix}$,
 because it appears through loop corrections of fermions having both $U(1)_Y$ and $U(1)_\chi$ charges
 even if $g_{\rm mix}$ vanishes at some scale.
In this paper, we impose $g_{\rm mix}(M_{\rm Pl})=0$,
 which would arise from breaking a simple unified gauge group into $SU(3)_c \otimes SU(2)_L \otimes U(1)_Y \otimes U(1)_\chi$.
In particular, there is the well-known
 decomposition of the $SO(10)$ GUT as $SO(10) \rightarrow SU(5) \otimes U(1)_\chi$.
Thus, when the $SO(10)$ GUT is broken at the Planck scale,
 $g_{\rm mix}(M_{\rm Pl})=0$ is naturally expected.

\begin{table}[t]
\begin{center}
\begin{tabular}{|c|cc|}\hline
 & $SU(3)_c \otimes SU(2)_L \otimes U(1)_Y$ & $U(1)_\chi$ \\
\hline \hline
$Q$ & (3, 2, 1/6) & 1/5\\
$U^c$ & ($\overline{3}$, 1, $-2/3$) & 1/5\\
$D^c$ & ($\overline{3}$, 1, 1/3) & $-3/5$\\
$L$ & (1, 2, $-1/2$) & $-3/5$\\
$E^c$ & (1, 1, 1) & 1/5\\
$N^c$ & (1, 1, 0) & 1\\
$H$ & (1, 2, 1/2) & $-2/5$\\
$\Phi$ & (1, 1, 0) & $2$\\
\hline
\end{tabular}
\caption{Quantum numbers of the fields in the SM with $U(1)_\chi$ symmetry.}
\label{particles}
\end{center}
\end{table}

Let us explain the mechanism of the $U(1)_\chi$ symmetry breaking
 and the subsequent EW symmetry breaking.
The $U(1)_\chi$ symmetry breaking is caused by the one-loop CW potential for the $U(1)_\chi$ sector,
 which is given by
\begin{eqnarray}
	V_\Phi(\phi) = \frac{1}{4} \lambda_\Phi \phi^4 + \frac{\phi^4}{64 \pi^2}
		\left( 10 \lambda_\Phi^2 + 48 g_\chi^4 - 8 \sum_{i=1}^3 y_{M_i}^4 \right)
		\left( \ln \frac{\phi^2}{M^2} - \frac{25}{6} \right),
\label{Vphi}
\end{eqnarray}
 around $\phi=M$ \cite{Coleman:1973jx}.
In this equation, we take $\Phi=\phi/\sqrt{2}$ in the unitary gauge,
 and Majorana Yukawa couplings of the right-handed neutrinos are diagonal
 as $Y_M^{ij}=y_{M_i} \delta_{ij}$.
In our following analyses,
 we will take $\sum y_{M_i}^4=N_\nu y_M^4$ for simplicity,
 where $N_\nu$ stands for the number of large Majorana Yukawa couplings
 that are enough to be effective in the RGE.
Equation (\ref{Vphi}) satisfies the following renormalization conditions
\begin{eqnarray}
	\left. \frac{\partial^2 V_\Phi}{\partial \phi^2} \right|_{\phi=0} = 0,\qquad
	\left. \frac{\partial^4 V_\Phi}{\partial \phi^4} \right|_{\phi=M} = 6 \lambda_\Phi.
\end{eqnarray}
When the SM singlet scalar has a nonzero VEV $\langle \phi \rangle = v_\Phi$,
 we choose the renormalization scale at $M = v_\Phi$
 to avoid the large log corrections, which have uncertainty in a large $\ln (\phi^2 / v_\Phi^2)$ region.
Then, the minimization condition of the potential (\ref{Vphi}) induces
\begin{eqnarray}
	\lambda_\Phi (v_\Phi) = \frac{11}{48 \pi^2}
		\left( 10 \lambda_\Phi^2 + 48 g_\chi^4 - 8 N_\nu y_M^4 \right) (v_\Phi).
\label{CW_relation}
\end{eqnarray}
When this relation is satisfied,
 the $U(1)_\chi$ symmetry is broken at $v_\Phi$.

Once the SM singlet scalar gets a nonzero VEV $v_\Phi$,
 the singlet scalar, the $Z'$ boson, and the right-handed neutrinos become massive:
\begin{eqnarray}
	M_\phi = \sqrt{\frac{6}{11} \lambda_\Phi (v_\Phi)} v_\Phi,\qquad
	M_{Z'} = 2 g_\chi(v_\Phi) v_\Phi,\qquad
	M_N = \sqrt{2} y_M(v_\Phi) v_\Phi,
\label{mass}
\end{eqnarray}
 respectively.
To realize the CW mechanism successfully,
 the logarithmic terms of potential (\ref{Vphi}) should be effective compared to the first term.
Thus, $\lambda_\Phi(v_\Phi)$ should be much smaller than $g_\chi(v_\Phi)$ and $y_M(v_\Phi)$,
 and the mass hierarchy $M_\phi \ll M_{Z'}$, $M_N$ is expected.
As will be shown later,
 the typical value of $M_\phi$ is a few GeV,
 and then the singlet scalar does not decouple in the EW scale,
 while the $Z'$ boson and the right-handed neutrinos decouple.
From Eq.~(\ref{CW_relation}), the masses are approximately written as
\begin{eqnarray}
	M_\phi^2 \approx \beta_{\lambda_\Phi} (v_\Phi) v_\Phi^2 >0, \qquad
	\frac{M_{Z'}}{M_N} \approx \left( \frac{2 N_\nu}{3} \right)^{1/4}.
\label{mass_approx}
\end{eqnarray}
Notice that $\beta_{\lambda_\Phi} (v_\Phi)>0$ is required,
 since the scalar mass squared must be positive.
On the other hand,
 $\beta_{\lambda_\Phi} (M_{\rm Pl})\leq0$ must be satisfied
 to avoid $\lambda_\Phi<0$ (which might cause the vacuum instability),
 since we impose $\lambda_\Phi(M_{\rm Pl})=0$.
Therefore, a running of $\lambda_\Phi$ is typically curved upward in the flatland scenario.

In general,
 a criterion for the successful CW mechanism has been derived as \cite{Hashimoto:2014ela}
\begin{eqnarray}
	K=\frac{123x^2-50x+12}{2+N_\nu} \sqrt{\frac{N_\nu}{6}} < 1,
\end{eqnarray}
 where $x$ represents a generalized $B-L$ gauge charge:
 $x=0$, 1/3, and $x=1/5$ correspond to $U(1)_R$, $U(1)_{B-L}$, and $U(1)_\chi$ models, respectively.
In our case, i.e., for a $U(1)_\chi$ model,
\begin{eqnarray}
	K=0.9417,\ 0.9988,\ 0.9786\quad {\rm for}\ N_\nu=1,\ 2,\ 3,
\end{eqnarray}
 respectively.
Thus, in the $U(1)_\chi$ model, the flatland scenario can work for any $N_\nu=1$--3.
However, in the $U(1)_R$ and $U(1)_{B-L}$ models,
 the flatland scenario cannot work because of $K>1$ for $N_\nu<10$ and 20, respectively. 
For $N_\nu=0$,
 $\lambda_\Phi$ becomes negative in any energy scale below the Planck scale,
 because $\beta_{\lambda_\Phi}$ almost depends on the gauge quartic terms (see Eq.~(\ref{beta_lphi})).
Thus, the flatland scenario cannot work in the $N_\nu=0$ case.

Here, we comment on a running of $\lambda_\Phi$ in the $N_\nu=2$ case,
 in which the value of $K$ is almost equal to 1.
It means that
 the terms $48 g_\chi^4 - 8 N_\nu y_M^4$ in $\beta_{\lambda_\Phi}$, Eq.~(\ref{beta_lphi}), are almost vanishing.
Then, two-loop order terms of $\beta_{\lambda_\Phi}$ are comparable to one-loop order terms,
 and $\beta_{\lambda_\Phi}$ becomes negative at all energy scales.
Thus, the running of $\lambda_\Phi$ is monotonically and very slowly decreasing
 from the EW scale to the Planck scale [see Fig.~\ref{quartic}-(b)],
 which is a quite different situation from that typically expected in the conventional flatland scenario.
It is worth noting that
 the CW mechanism can also work in the $N_\nu=2$ case,
 since the minimization condition (\ref{CW_relation}) can be satisfied at the energy scale of $v_\Phi$.

After the $U(1)_\chi$ symmetry breaking by the CW mechanism,
 the Higgs mass term is generated as
\begin{eqnarray}
	m_H^2 (v_\Phi) = \frac{1}{2} \lambda_{\rm mix} (v_\Phi) v_\Phi^2,
\end{eqnarray}
 and the tree-level Higgs potential at $v_\Phi$ is given by
\begin{eqnarray}
	V_H(h) = \frac{1}{4} \lambda_H (v_\Phi) h^4 + \frac{1}{2} m_H^2 (v_\Phi) h^2,
\end{eqnarray}
 where we take $H = (0,\ (v_H + h)/\sqrt{2})^T$ in the unitary gauge.
Below the energy scale of $v_\Phi$,
 running of the Higgs mass term is governed by
\begin{eqnarray}
	\frac{d m_H^2}{d \ln \mu} = \frac{1}{16\pi^2}
		\left[ m_H^2 \left( 12 \lambda_H + 6 y_t^2 -\frac{9}{2} g_2^2 - \frac{3}{2} g_Y^2 -\frac{3}{2} \left( g_{\rm mix} - \frac{4}{5} g_\chi \right)^2 \right) + 2 \lambda_{\rm mix} M_\phi^2 \right].
\label{beta_mH}
\end{eqnarray}
From Eq.~(\ref{mass}),
 the last term in Eq.~(\ref{beta_mH}) is of the order of $\lambda_\Phi m_H^2$,
 and then it is negligible because of $\lambda_\Phi \ll 1$.
In other words,
 $\lambda_{\rm mix} \sim (v_H/v_\Phi)^2$ is required since $m_H^2$ is the EW scale,
 and then it is small enough to be neglected.
Below $M_{Z'}$, the $Z'$ boson decouples,
 and then the terms including $g_{\rm mix}$ and/or $g_\chi$ are omitted from Eq.~(\ref{beta_mH}).
Note that the effects can be numerically neglected,
 since they are sufficiently small compared to other contributions in Eq.~(\ref{beta_mH}).
As the VEV of the Higgs $v_H$, the minimization condition of the Higgs potential induces
\begin{eqnarray}
	v_H = \sqrt{ \frac{- m_H^2 (v_H)}{\lambda_H (v_H)} },
\end{eqnarray}
 where $m_H^2$ must be negative to realize the electroweak symmetry breaking.
Notice that
 $\lambda_{\rm mix}$, or $m_H^2$, naturally becomes negative in the flatland scenario,
 since $\beta_{\lambda_{\rm mix}}$ strongly depends on the gauge quartic terms
 which are always positive [see Eq.~(\ref{beta_lmix})].
Then, the Higgs pole mass is given by
\begin{eqnarray}
	M_h^2 = 2 \lambda_H (v_H) v_H^2 + \Delta M_h^2,
\label{Mh}
\end{eqnarray}
 where $\Delta M_h^2$ is the Higgs self-energy correction to the Higgs pole mass
 \cite{Degrassi:2012ry}.
The running of couplings controlled by the initial values of $g_\chi$ and $y_M$,
 and they are determined to realize $v_H \simeq 246$\,GeV and $M_h \simeq 125$\,GeV.
On the other hand, once $g_\chi$ or $y_M$ is fixed,
 the other is uniquely determined by Eq.~(\ref{CW_relation}).
Therefore, there is only one free parameter in the flatland scenario,
 and the physical quantities are uniquely predicted.\footnote{
More accurately,
 there are more degrees of freedom for the Majorana Yukawa coupling matrix $Y_M^{ij}$.
But, we had taken ${\rm tr} [Y_M^{ij}]= N_\nu y_M$ for simplicity,
 and analyze independently by fixing $N_\nu=1$, 2, and 3.}

After the EW symmetry breaking,
 the singlet scalar and the Higgs are mixed by the $\lambda_{\rm mix}$ term.
Then, the mass eigenvalues are different from $M_\phi$ and $M_h$.
The scalar mass squared matrix is given by
\begin{eqnarray}
	{\cal M}^2 = \left( 
			\begin{array}{cc}
				M_h^2 & \frac{1}{2} \lambda_{\rm mix} v_H v_\Phi \\
				\frac{1}{2} \lambda_{\rm mix} v_H v_\Phi & M_\phi^2
			\end{array} \right),
\label{mass_matrix}
\end{eqnarray}
 where $M_h$ and $M_\phi$ are given by Eqs.~(\ref{Mh}) and (\ref{mass}), respectively.
Then, the scalar mixing angle $\theta$ is expressed as
\begin{eqnarray}
	\tan 2\theta = \frac{\lambda_{\rm mix} v_H v_\Phi}{M_h^2 - M_\phi^2}.
\end{eqnarray}
Since the flatland scenario expects $\lambda_\Phi \ll |\lambda_{\rm mix}| \ll \lambda_H$
 at a low energy scale,
 the lighter scalar mass squared eigenvalue is approximately written by
\begin{eqnarray}
 	M_{\phi'}^2 
				&\approx& M_\phi^2
				- \frac{\lambda_{\rm mix}^2 v_H^2 v_\Phi^2}{4\left( M_h^2 - M_\phi^2 \right)}.
\label{light_mass}
\end{eqnarray}
It would be negative for a large $|\lambda_{\rm mix}|$.
We will discuss the positive definiteness of the scalar mass squared eigenvalues
 in the next section.

In the same way, the $U(1)$ gauge bosons are mixed by the $g_{\rm mix}$ term.
It is potentially dangerous, because the $\rho$-parameter deviates from unity at the tree level.
The mass term of the $Z$ and $Z'$ bosons are given by
\begin{eqnarray}
	{\cal L}_Z = \frac{1}{2}(Z_\mu, {Z'}_\mu) M_{ZZ'}^2
		\left( \begin{array}{c} Z^\mu\\ {Z'}^\mu \end{array} \right),\
	M_{ZZ'}^2 = \left( \begin{array}{cc}
		M_Z^2 & \delta M^2 \\
		\delta M^2 & M_{Z'}^2 + \frac{1}{4}\left( g_{\rm mix}- \frac{4}{5}g_\chi \right)^2 v_H^2
						\end{array} \right),
\end{eqnarray}
 where $M_Z$ is the SM one as $M_Z^2 = (g_Y^2+g_2^2)v_H^2/4$,
 and the second term of $Z'$ boson mass is obtained by the Higgs VEV after the EW symmetry breaking,
 which is much smaller than $M_{Z'}^2$ because of $v_H \ll v_\Phi$.
The mixing term is given by
\begin{eqnarray}
	\delta M^2 = \frac{1}{4} \sqrt{g_Y^2+g_2^2} \left( g_{\rm mix} - \frac{4}{5} g_\chi \right) v_H^2,
\end{eqnarray}
 and the mass matrix is diagonalized by
\begin{eqnarray}
	\tan 2\theta_Z = \frac{2\delta M^2}{M_Z^2 - \left( M_{Z'}^2 + \frac{1}{4}\left( g_{\rm mix}- \frac{4}{5}g_\chi \right)^2 v_H^2 \right)}.
\end{eqnarray}
After diagonalizing the mass matrix,
 the lighter mass squared eigenvalue is approximately obtained by
\begin{eqnarray}
	M_{1}^2 \approx M_Z^2 - \frac{\left( \delta M^2 \right)^2}{\left( M_{Z'}^2 + \frac{1}{4}\left( g_{\rm mix}- \frac{4}{5}g_\chi \right)^2 v_H^2 \right) -M_Z^2 },
\label{M1}
\end{eqnarray}
 which is smaller than $M_Z^2$.
The $\rho$-parameter deviates from unity when $M_1$ is different from $M_Z$.
We will also discuss the deviation of the $\rho$-parameter in the next section.

\section{Constraints by the vacuum stability} \label{sec:vacuum}

\begin{figure}[t]
  \begin{center}
	\flushleft \hspace{3.5cm} (a) $N_\nu=1$ \hspace{6cm} (b) $N_\nu=2$ \\
          \includegraphics[clip, height=5.2cm]{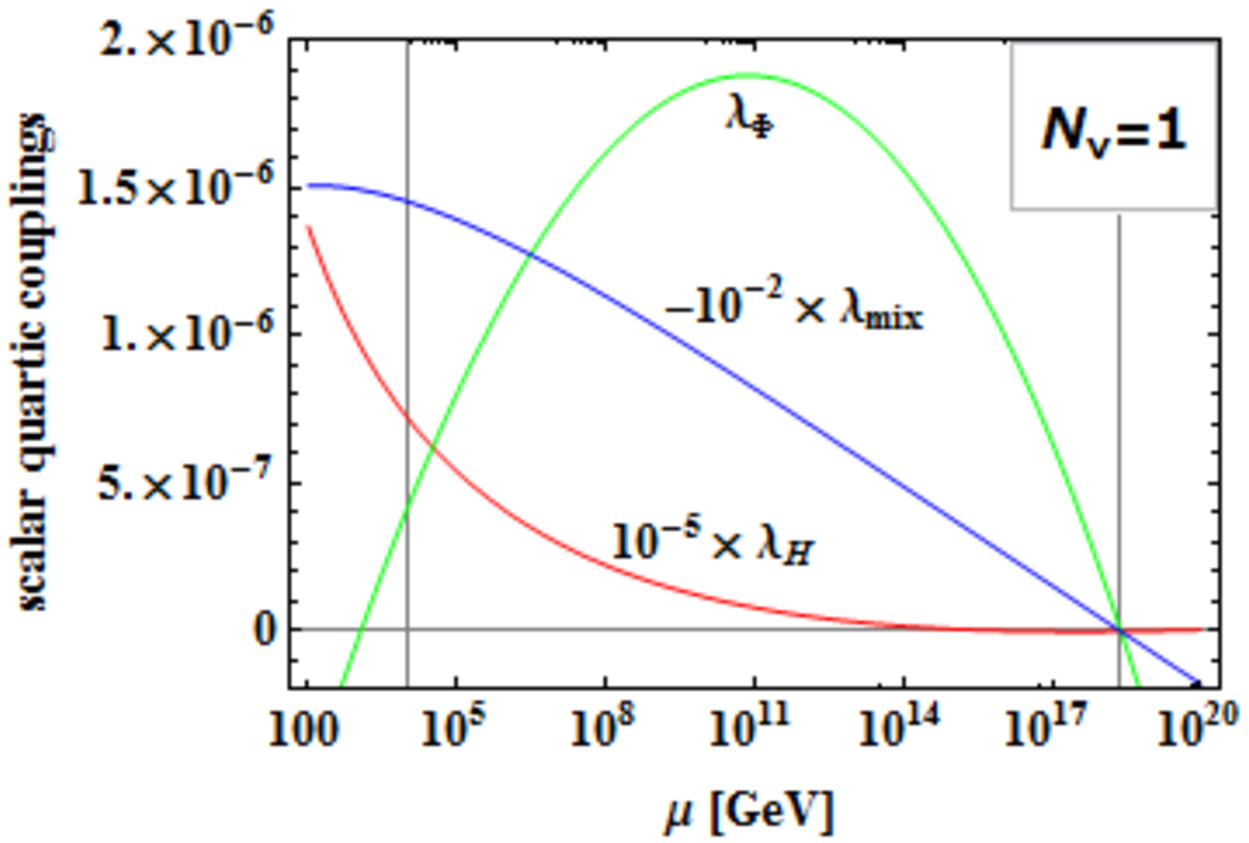} \hspace{0.5cm}
          \includegraphics[clip, height=5.2cm]{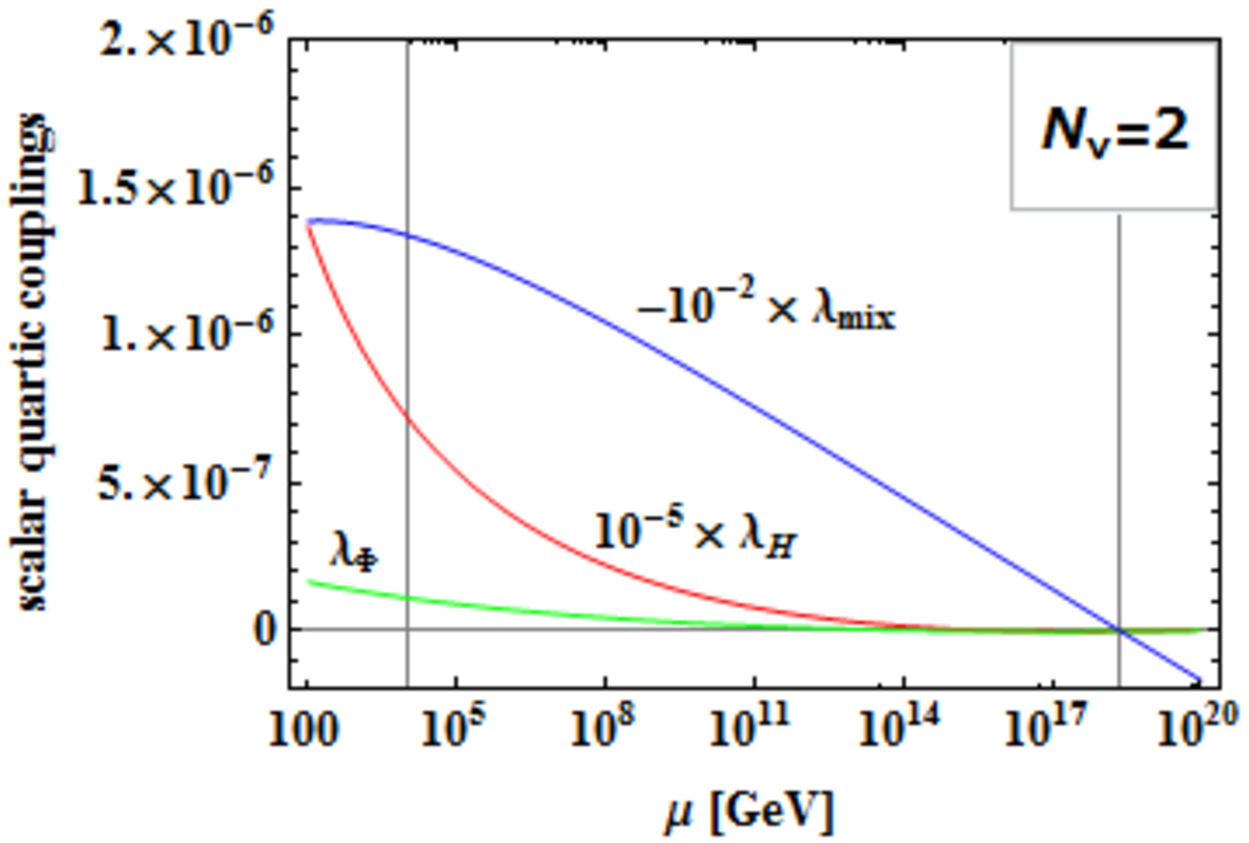} \\
	\flushleft \hspace{3.5cm} (c) $N_\nu=3$ \\
          \includegraphics[clip, height=5.2cm]{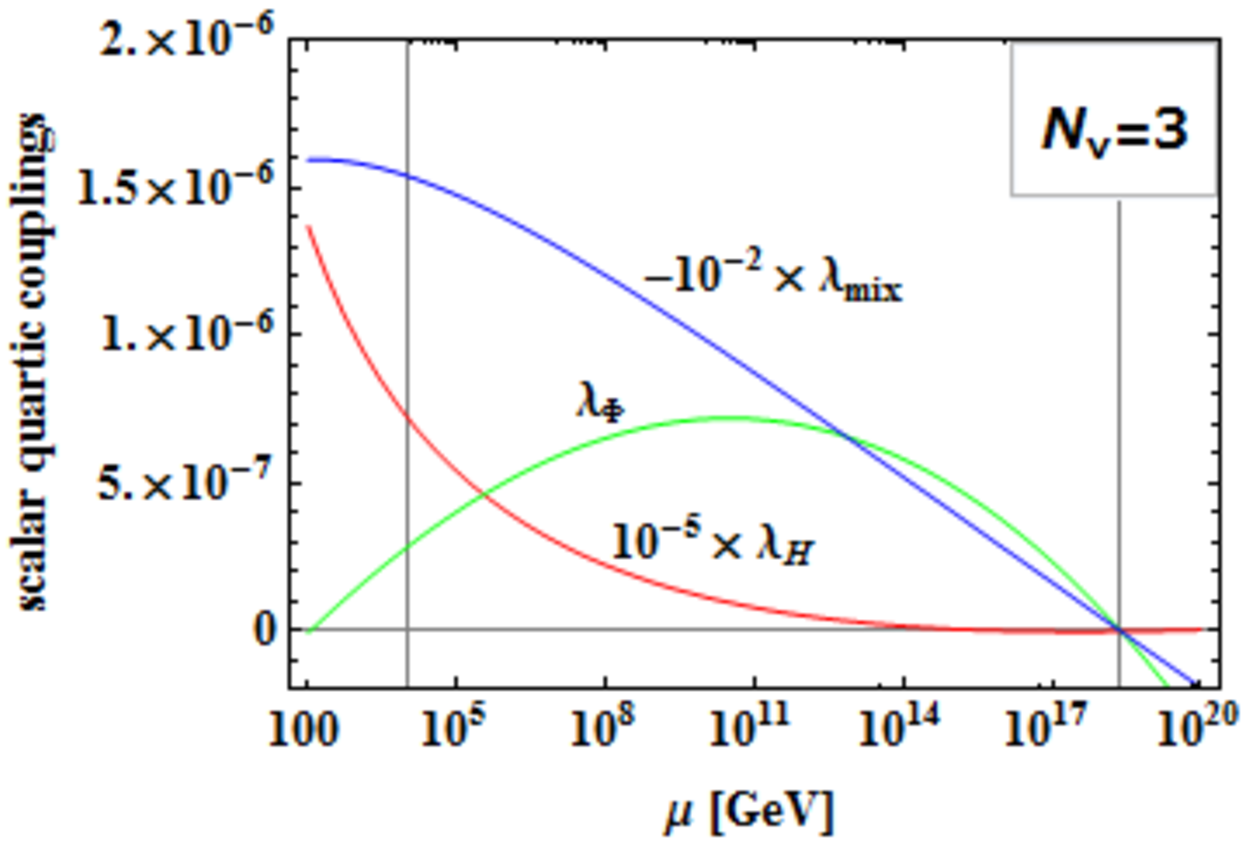}
  \end{center}
\caption{Example of runnings of quartic couplings for $N_\nu=1$, 2 and 3.
The red, green, and blue lines correspond to $10^{-5} \times \lambda_H$,
 $\lambda_\Phi$, and $-10^{-2} \times \lambda_{\rm mix}$, respectively.
Two vertical grid lines represent $v_\Phi$ and $M_{\rm Pl}$, respectively.
The decoupling effects of the $Z'$ boson and the right-handed neutrinos are not considered
 in these figures.}
\label{quartic}
\end{figure}

In Fig.~\ref{quartic},
 we show runnings of the scalar quartic couplings.
The $N_\nu=1$ and 3 cases show the behavior as expected in the conventional flatland scenario,
 that is, a running of $\lambda_\Phi$ is curved upward,
 and $\lambda_{\rm mix}$ is negative to realize the negative Higgs mass term.
On the other hand, for $N_\nu =2$,
 a running of $\lambda_\Phi$ behaves quite differently from $N_\nu=1$ and 3.
The running of $\lambda_\Phi$ is monotonically and slowly decreasing
 from the EW scale to the Planck scale as mentioned above.
In all $N_\nu=1$--3,
 the Higgs mass of 125.1\,GeV is realized with the top pole mass of 171\,GeV,
 and the $U(1)_\chi$ is broken at $v_\Phi \simeq 10$\,TeV.
In these cases, the singlet scalar, the $Z'$ boson, and right-handed neutrino
 are massive for $N_\nu= 1$, 2, 3 as
 $M_\Phi \simeq 5.0$\,GeV, 2.7\,GeV, 3.9\,GeV, $M_{Z'} \simeq 2.0$\,TeV, 2.0\,TeV, 2.0\,TeV,
 and $M_N \simeq 2.3$\,TeV, 1.9\,TeV, 1.7\,TeV,
 respectively.
The values of the ratio $M_{Z'}/M_N$ agree very well with the predicted values from Eq.~(\ref{mass_approx}).

We investigate the parameter spaces allowed by the vacuum stability
 using two-loop RGEs.
Since there is a few percent error for a running of the Higgs quartic coupling $\lambda_H$
 in one-loop RGEs,
 we have to use two-loop RGEs for a discussion of the vacuum stability.
Adding the singlet scalar into the SM,
 the vacuum stability conditions are given by \cite{Chakrabortty:2013zja}
\begin{eqnarray}
	\lambda_H>0,\qquad \lambda_\Phi>0,\qquad 4\lambda_H \lambda_\Phi - \lambda_{\rm mix}^2>0.
\label{conditions}
\end{eqnarray}
These conditions should be satisfied in any energy scale.
If all the quartic couplings are positive,
 the potential is trivially bounded from below,
 and the vacuum is stable.
The last condition in Eq.~(\ref{conditions}) shows
 the upper bound of $|\lambda_{\rm mix}|$.
Note that there are the non-trivial vacuum stability conditions of $\lambda_{\rm mix}<0$.

\begin{figure}[t]
  \begin{center}
          \includegraphics[clip, scale=0.8]{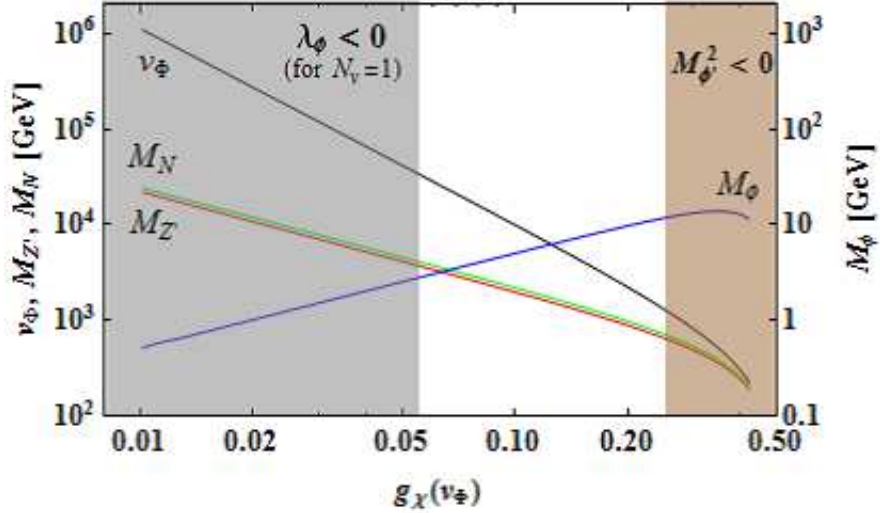}
  \end{center}
\caption{$U(1)_\chi$ gauge coupling dependences on 
 the singlet scalar VEV and new particle masses obtained by Eq.~(\ref{mass}).
The left and right shaded regions are excluded by the $\lambda_\Phi<0$
 and $M_{\phi'}^2<0$ conditions, respectively.
This figure shows the $N_\nu=1$ case,
 and the left shaded region does not appear in the $N_\nu=2$ and 3 cases.
}
\label{result}
\end{figure}

For our analyses,
 we take $g_\chi$ as a free parameter,
 and show its dependences on the other physical quantities in Fig.~\ref{result}.
Since $M_{Z'}$ and $M_N$ satisfy Eq.~(\ref{mass_approx}),
 they are almost the same value.
Although this figure shows the result for $N_\nu=1$,
 the predicted physical quantities are almost the same for $N_\nu=2$ and 3.
This is because the runnings of the couplings, except $\lambda_\Phi$,
 are almost the same for any $N_\nu$.
The left and right shaded regions correspond to constraints obtained
 by the vacuum stability conditions
 and the positive definiteness of the scalar mass squared eigenvalues, respectively.
We will explain the constraints while discussing each condition below.

\begin{figure}[t]
  \begin{center}
          \includegraphics[clip, height=5cm]{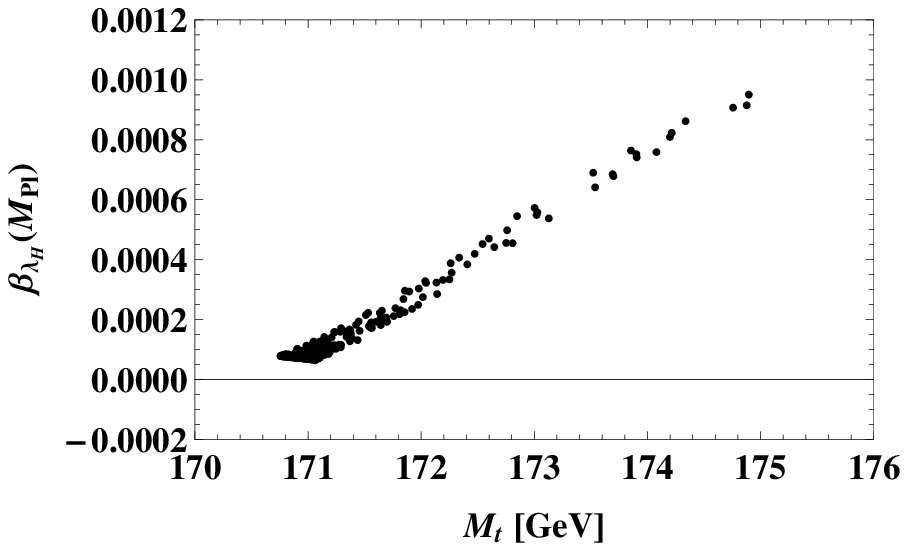} \hspace{0.2cm}
          \includegraphics[clip, height=5cm]{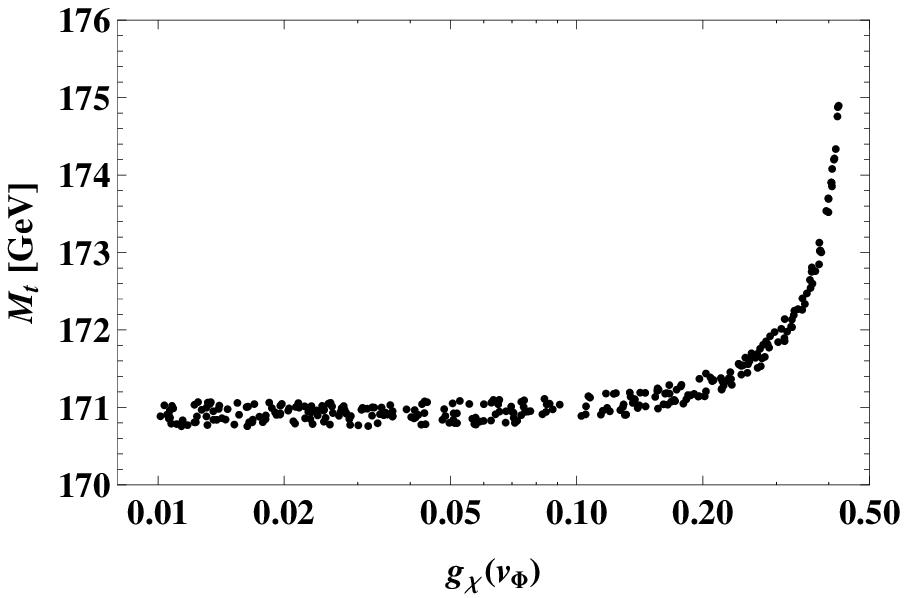}
  \end{center}
\caption{Left: Relation between the top pole mass
 and the $\beta$ function of $\lambda_H$ at the Planck scale.
 Right: $g_\chi$ dependences on the top pole mass.
 The dots realize the Higgs mass in a range of Eq.~(\ref{Higgs}).
}
\label{mt-beta}
\end{figure}

First, we consider the Higgs quartic coupling $\lambda_H$.
To realize $\lambda_H>0$ in any energy scale,
 the $\beta$ function of $\lambda_H$ at the Planck scale
 should satisfy $\beta_{\lambda_H}(M_{\rm Pl})\leq0$ because of $\lambda_H(M_{\rm Pl})=0$.
In the SM,
 once $\lambda_H(M_{\rm Pl})=0$ and $\beta_{\lambda_H}(M_{\rm Pl})\leq0$ is imposed,
 we can find $M_t \gtrsim 173$\,GeV and $M_h \gtrsim 129$\,GeV
 \cite{Buttazzo:2013uya,Haba:2014oxa},
 while this lower bound of the Higgs mass is disfavored by the experiments.
In the flatland scenario, $\beta_{\lambda_H}(M_{\rm Pl})$ is given by
\begin{eqnarray}
	\beta_{\lambda_H} (M_{\rm Pl}) = \frac{1}{(4\pi)^2} \left[ 
		- 6 y_t^4 + \frac{3}{8} \left\{ 2 g_2^4 + \left( g_2^2 + g_Y^2
		+ \frac{16}{25} g_\chi^2 \right)^2 \right\} \right],
\label{beta_lambda_H}
\end{eqnarray}
 up to the one-loop level.
The larger $g_\chi$ becomes,
 the larger the top Yukawa coupling $y_t$ (or the top pole mass $M_t$) becomes compared with the SM
 in order to realize the 125\,GeV Higgs mass.
The left figure of Fig.~\ref{mt-beta} shows
 the relation between $M_t$ and $\beta_{\lambda_H}(M_{\rm Pl})$,
 in which the dots realize the Higgs mass in the range of Eq.~(\ref{Higgs}).
Then, the larger $M_t$ becomes, the larger $\beta_{\lambda_H}(M_{\rm Pl})$ becomes,
 while the Higgs mass cannot be realized by $M_t\lesssim 171$\,GeV.
We find that
 it is impossible to simultaneously realize both
 $\beta_{\lambda_H}(M_{\rm Pl})\leq0$ (or $\lambda_H>0$)
 and $M_h \simeq 125$\,GeV.

On the other hand,
 once one gives up $\lambda_H>0$ in any energy scale
 and imposes $\lambda_H(M_{\rm Pl})=0$,
 the measured Higgs mass as $M_h \simeq 125$\,GeV can be realized
 by $M_t \simeq 171$\,GeV in the SM.
Although $\lambda_H$ becomes negative below the Planck scale,
 the vacuum is meta-stable, which is phenomenologically allowed.
The same thing can be said in the flatland scenario
 unless the running of $\lambda_H$ does not drastically change from that in the SM.
As $g_\chi$ becomes larger,
 $M_h \simeq 125$\,GeV can be realized by the larger $M_t$ compared to the SM case,
 which is shown in the right figure of Fig.~\ref{mt-beta}.
When we allow $\lambda_H<0$ as long as the vacuum is meta-stable,
 $M_h \simeq 125$\,GeV can be realized by $g_\chi \simeq 0.4$
 corresponding to the experimentally favored value, $M_t\simeq 173$\,GeV \cite{ATLAS:2014wva}.
However, the large $g_\chi$ region as $g_\chi \gtrsim 0.2$ is excluded for $N_\nu=1$
 by the positive definiteness of the scalar mass squared eigenvalues,
 as mentioned below.

Next, we consider the singlet scalar quartic coupling $\lambda_\Phi$.
In Fig.~\ref{quartic}~(a),
 $\lambda_\Phi$ seems to become negative an order of magnitude below the singlet scalar VEV $v_\Phi$.
However, in fact, we can find $\lambda_\Phi>0$ is realized as follows.
After the $U(1)_\chi$ symmetry breaking,
 the $Z'$ boson and the right-handed neutrinos become massive.
Since their masses are the same order of magnitude as $v_\Phi$,
 they would decouple and be integrated out from the theory
 before $\lambda_\Phi$ becomes negative.
Then, the $\beta$ function of $\lambda_\Phi$ becomes
\begin{eqnarray}
	\beta_{\lambda_\Phi} (\mu<M_{Z'},M_N) &=& \frac{1}{(4\pi^2)}
		\left[ 20 \lambda_\Phi^2 + 2 \lambda_{\rm mix}^2 \right],
\end{eqnarray}
 up to the one-loop level.
It does not include contributions of loop diagrams
 which have internal lines of the $Z'$ boson and/or the right-handed neutrinos.
Since both $\lambda_\Phi$ and $\lambda_{\rm mix}$ are numerically almost equal to zero around $v_\Phi$,
 i.e., $\beta_{\lambda_\Phi} (\mu<M_{Z'},M_N) \simeq 0$,
 it is reasonable to consider
 $\lambda_\Phi(\mu<M_{Z'},M_N) \simeq \lambda_\Phi(M_{Z'}) \simeq \lambda_\Phi(M_N)$.\footnote{
Here, we consider the tree-level matching condition,
 that is, the running couplings have no gaps at $M_{Z'}$ and $M_N$.
}
Thus, we can find that
 the parameter space of $g_\chi (\simeq y_M) \lesssim 0.055$ is excluded
 by $\lambda_\Phi (\mu<M_{Z'},M_N)<0$,
 which is shown as the left shaded region in Fig.~\ref{result}.
This constraint corresponds to
 $v_\Phi \lesssim 3.3 \times 10^5\,{\rm GeV}$,
 $M_\Phi \gtrsim 2.8\,{\rm GeV}$,
 $M_{Z'} \lesssim 3.7\,{\rm TeV}$, and
 $M_N \lesssim 4.1\,{\rm TeV}$, respectively.

As for $N_\nu=2$ and 3,
 we find that $\lambda_\phi>0$ is not a constrained condition.
For $N_\nu=2$,
 we required that the running of $\lambda_\Phi$ is monotonically decreasing
 from the EW scale to the Planck scale,
 as in Fig.~\ref{quartic}~(b).
Since $\lambda_\Phi$ becomes rather larger at lower energy scales,
 $\lambda_\Phi$ is positive at any energy scale.
Thus, the condition $\lambda_\phi>0$ gives no constraint for $N_\nu=2$.
For $N_\nu=3$, the running of $\lambda_\Phi$ is the similar to that for $N_\nu=1$,
 but the gradient of the running is much gentler,
 as in Fig.~\ref{quartic}~(c).
Then, even for $g_\chi \sim 0.01$
 the $Z'$ boson and the right-handed neutrinos are decoupled
 before $\lambda_\Phi$ becomes negative.
Therefore, the small $g_\chi$ regions are almost not constrained for $N_\nu=3$.

\begin{figure}[t]
  \begin{center}
          \includegraphics[clip, height=5.3cm]{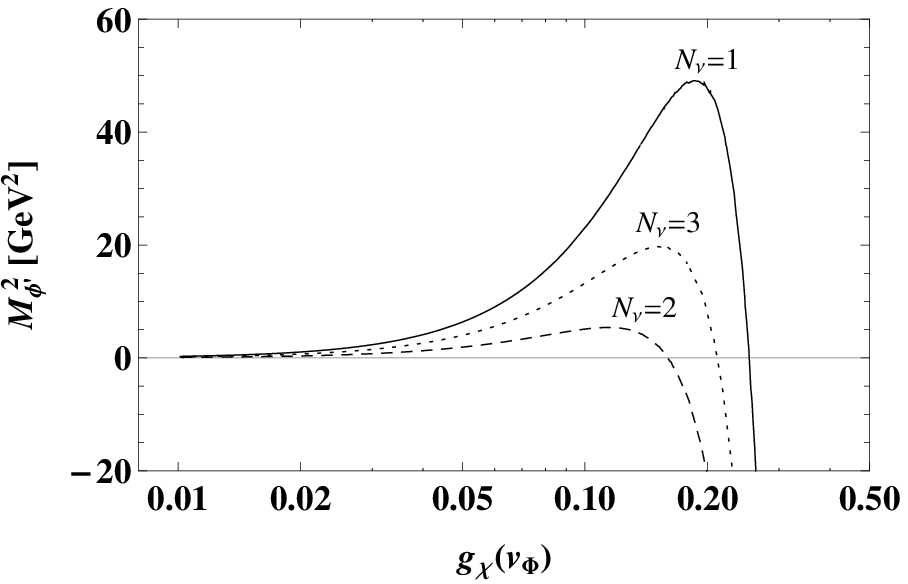}
          \includegraphics[clip, height=5.3cm]{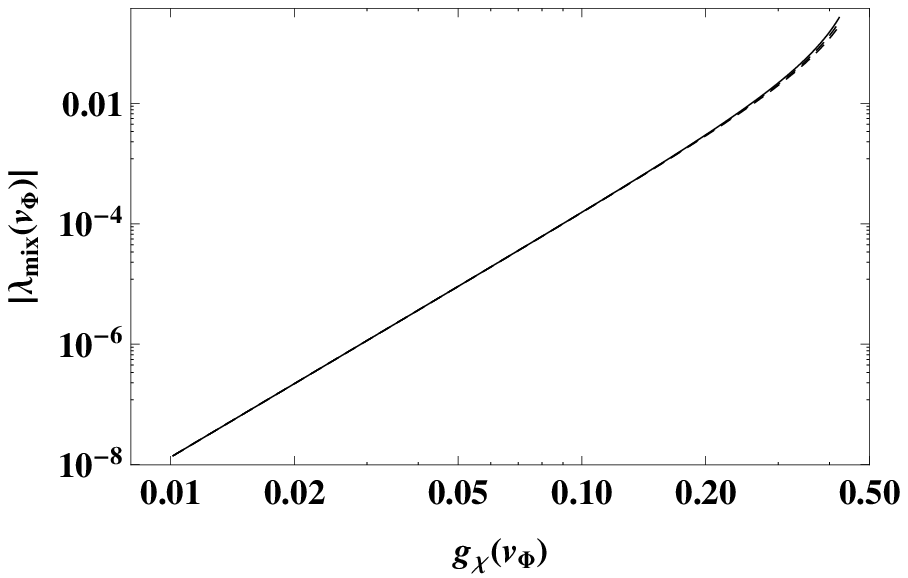}
  \end{center}
\caption{$g_\chi$ dependences on the lighter scalar mass squared eigenvalue (left)
 and the scalar mixing coupling (right).
The solid, dashed, dotted lines correspond to $N_\nu=1$, 2, and 3 respectively.
}
\label{mphi}
\end{figure}

Next, we consider the mixing coupling between the scalar fields $\lambda_{\rm mix}$.
The vacuum stability requires $4\lambda_H \lambda_\Phi - \lambda_{\rm mix}^2>0$,
 which means the large mixing can be excluded.
When both $\lambda_H$ and $\lambda_\Phi$ are positive,
 the inequality is almost always satisfied
 because of $\lambda_H \gg |\lambda_{\rm mix}|$.
On the other hand, the inequality cannot be explicitly satisfied
 when either $\lambda_H$ or $\lambda_\Phi$ is negative.
Then, we can find that
 the condition $4\lambda_H \lambda_\Phi - \lambda_{\rm mix}^2>0$ is almost the same
 as the condition $\lambda_H>0$.
Note that
 $4\lambda_H \lambda_\Phi - \lambda_{\rm mix}^2>0$ cannot be satisfied in all energy scales,
 since $\lambda_H>0$ cannot be satisfied below the Planck scale
 in order to realize the Higgs mass of 125\,GeV as mentioned above.
Thus, we try to constrain $\lambda_{\rm mix}$ in other conditions,
 that is, the positive definiteness of the scalar mass squared eigenvalues.
The lighter scalar mass squared $M_{\phi'}^2$ given by Eq.~(\ref{light_mass})
 would be negative for a large $|\lambda_{\rm mix}|$.
The left figure of Fig.~\ref{mphi} shows that
 $M_{\phi'}^2$ becomes negative for the large $g_\chi$ region,
 which corresponds to a large mixing region (see the right figure).
Since the running of $\lambda_{\rm mix}$ is almost the same for any $N_\nu=1$--3,
 the relation between $g_\chi$ and $\lambda_{\rm mix}$ is also the same.
Thus, considering the positive definiteness of the scalar mass squared eigenvalues,
 we can find that
 large $g_\chi$ regions are excluded in $g_\chi \gtrsim 0.25$, 0.16, and 0.23
 for $N_\nu=1$, 2, and 3, respectively.
For example, in $N_\nu=1$ case, it is shown as the right shaded region in Fig.~\ref{result}.
This constraint corresponds to
 $v_\Phi \gtrsim 1.3\,{\rm TeV}$,
 $M_\phi \lesssim 12\,{\rm GeV}$,
 $M_{Z'} \gtrsim 650\,{\rm GeV}$, and
 $M_N \gtrsim 720\,{\rm GeV}$, respectively.
Therefore, the physical quantities are constrained from both above and below for $N_\nu=1$.
We show the allowed parameter regions for the physical quantities
 in Table \ref{table}.
In fact, the ATLAS and CMS experiments have obtained larger lower bounds for $M_{Z'}$
 than those in Table \ref{table}
 as mentioned below.

\begin{table}[t]
\begin{center}
\begin{tabular}{|c|ccc|}\hline
 & $N_\nu=1$ & $N_\nu=2$ & $N_\nu=3$ \\
\hline \hline
$g_\chi$ &
 $0.055 \lesssim g_\chi \lesssim 0.25$ &
 $g_\chi \lesssim 0.16$ &
 $g_\chi \lesssim 0.23$ \\
$v_\Phi$ &
 $1.3\,{\rm TeV} \lesssim v_\Phi \lesssim 3.3 \times 10^5\,{\rm GeV}$ &
 $3.8\,{\rm TeV} \lesssim v_\Phi$ &
 $2.0\,{\rm TeV} \lesssim v_\Phi$ \\
$M_\phi$ &
 $2.8\,{\rm GeV} \lesssim M_\phi \lesssim 12\,{\rm GeV}$ &
 $M_\phi \lesssim 4.2\,{\rm GeV}$&
 $M_\phi \lesssim 7.7\,{\rm GeV}$ \\
$M_{Z'}$ &
 $650\,{\rm GeV} \lesssim M_{Z'} \lesssim 3.7\,{\rm TeV}$ &
 $1.2\,{\rm TeV} \lesssim M_{Z'}$ &
 $860\,{\rm GeV} \lesssim M_{Z'}$ \\
$M_N$ &
 $720\,{\rm GeV} \lesssim M_N \lesssim 4.1\,{\rm TeV}$ &
 $1.1\,{\rm TeV} \lesssim M_N$ &
 $720\,{\rm GeV} \lesssim M_N$ \\
\hline
\end{tabular}
\caption{Allowed parameter regions for the physical quantities.
}
\label{table}
\end{center}
\end{table}

\section{Experimental bounds} \label{sec:ex-bound}

In this section, we mention the experimental bounds.
When there is gauge mixing between the $Z$ and $Z'$ bosons in the EW scale,
 it is dangerous since the $\rho$-parameter deviates from unity at the tree level.
Let us estimate the deviation of the $\rho$-parameter \cite{Hashimoto:2014ela}.
The tree-level $\rho$-parameter is defined by $\rho_0 = M_W^2/(M_1^2 c_W^2)$,
 where $M_W^2=g_2^2 v_H^2/4$ is the $W$ boson mass,
  and $c_W^2=g_2/\sqrt{g_Y^2+g_2^2}$ is the Weinberg angle.
The deviation of the $\rho$-parameter $\delta \rho \equiv \rho_0 - 1$ is always positive
 because of $M_1 < M_Z$.
From Eq.~(\ref{M1}), $\delta \rho$ is approximately given by
\begin{eqnarray}
	\delta \rho \equiv \rho_0 - 1 
		\approx \frac{v_H^2}{4 \left[ \left( M_{Z'}^2 + \frac{1}{4}\left( g_{\rm mix}- \frac{4}{5}g_\chi \right)^2 v_H^2 \right) -M_Z^2 \right]}
			\left( g_{\rm mix} - \frac{4}{5} g_\chi \right)^2.
\label{delrho}
\end{eqnarray}
We can find that $\delta \rho$ is proportional to $\tan 2\theta_Z$.
Thus, $\delta \rho$ is vanishing in the limit of $\tan 2\theta_Z \rightarrow 0$,
 which is necessarily required.

\begin{figure}[t]
  \begin{center}
          \includegraphics[clip, scale=1]{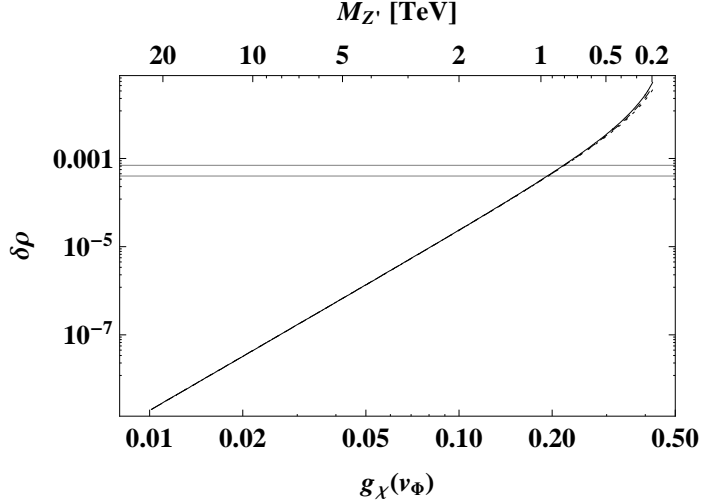}
  \end{center}
\caption{$g_\chi$ and $M_{Z'}$ dependence on $\delta \rho$.
The lower and upper horizontal lines correspond to
 the central value and the upper bound at 1 $\sigma$, respectively.
The solid, dashed, dotted lines correspond to $N_\nu=1$, 2, and 3 respectively.
}
\label{gx-drho}
\end{figure}

Now, we can compare $\delta \rho$ with its experimental bound
 $\rho_0=1.0004^{+0.0003}_{-0.0004}$ \cite{Beringer:1900zz}.
Figure \ref{gx-drho} shows $g_\chi$ and $M_{Z'}$ dependence on $\delta \rho$,
 in which the lower and upper horizontal lines correspond to
 the central value and the upper bound at 1 $\sigma$, respectively.
We can see that $\delta \rho$ is almost independent of $N_\nu$,
 since $N_\nu$ does not change the running of gauge couplings up to one-loop level.
$\delta \rho$ becomes larger as $g_\chi$ becomes larger, equivalently $M_{Z'}$ becomes lower.
Then, the central value of $\rho_0$ and its upper bound at 1 $\sigma$ correspond to
 $g_X \simeq 0.19$ and 0.21, equivalently $M_{Z'} \simeq 950$\,GeV and 820\,GeV, respectively.
Thus, $M_{Z'}$ should be heaver than 820\,GeV.

Finally, we mention the $Z'$ boson mass bounds obtained by the recent collider experiments
 (see Ref.~\cite{Moortgat-Picka:2015yla} for a review).
Currently, the highest mass bounds on the $Z'$ boson are obtained
 by searches at the LHC by the ATLAS and CMS experiments.
The most recent results are based on the search for the heavy neutral gauge boson decaying to
 $e^+ e^-$ or $\mu^+ \mu^-$ pairs.
The ATLAS obtains the exclusion limits at 95$\%$ C.L.
 as $M_{Z'} > 2.24$\,TeV for the $U(1)_\chi$ model.
It is used the center-of-mass energy $\sqrt{s}=8$\,TeV
 $pp$ collision data set collected in 2012
 corresponding to an integrated luminosity of approximately
 5.9 ($e^+ e^-$) / 6.1 ($\mu^+ \mu^-$) fb$^{-1}$ \cite{ATLAS:2012ipa}.
Similarly, the CMS obtains the exclusion limits at 95$\%$ C.L.
 as $M_{Z'} > 2.59$\,TeV for the sequential standard model with SM-like couplings
 \cite{Altarelli:1989ff}.
It used the $\sqrt{s}=8$\,TeV $pp$ collision data set
 and $\sqrt{s}=7$\,TeV data set collected by the CMS experiment in 2011
 corresponding to an integrated luminosities of up to 4.1 fb$^{-1}$ \cite{CMS:2012cba}.

In addition, another constraint is obtained by measurements of $e^+ e^- \rightarrow f \bar{f}$
 above the $Z$-pole at the LEP-II, where $f$ denotes various SM fermions.
When $M_{Z'}$ is larger than the largest collider energy of the LEP-II,
 which is about 209\,GeV,
 one can effectively perform an expansion in $s/M_{Z'}^2$ for four fermion-interactions.
Then, effective four-fermion interactions have been bounded by the LEP-II.
Since the amplitudes of the $Z'$ boson mediating interactions are proportional to $g_{Z'}^2/M_{Z'}^2$,
 the bound can be obtained as the ratio $M_{Z'}/g_{Z'}$,
 where $g_{Z'}$ is a flavor independent $Z'$ gauge coupling.
Using the single channel estimation,
 one can obtain the lower bound $M_{Z'}/g_{\chi} \gtrsim 3.8$\,TeV for the $U(1)_\chi$ model
 \cite{Carena:2004xs}.
In a recent parameter fitting analysis,
 the lower bound $M_{Z'}/g_{\chi} \geq 4.8$\,TeV has been obtained at 99$\%$ C.L.
 \cite{Cacciapaglia:2006pk}.

\begin{figure}[t]
  \begin{center}
          \includegraphics[clip, scale=1]{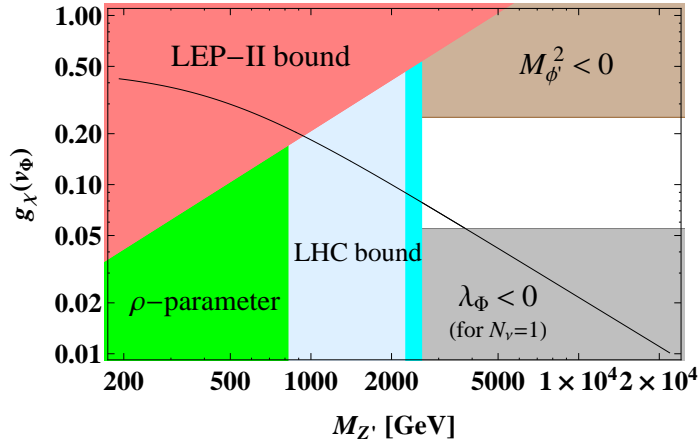}
  \end{center}
\caption{All the constraints on $M_{Z'}$ and $g_\chi$.
The black line corresponds to the flatland prediction for $N_\nu=1$.
The shaded regions show constraints obtained by Sect.~\ref{sec:vacuum} and \ref{sec:ex-bound}.
}
\label{all_const}
\end{figure}

Let us summarize all the constraints in Fig.~\ref{all_const}.
In the flatland scenario,
 the physical quantities are uniquely determined once one parameter is fixed.
The relation between $M_{Z'}$ and $g_\chi$ are given by the black solid line. 
The shaded regions show constraints obtained by Sects.~\ref{sec:vacuum} and \ref{sec:ex-bound}.
The constraint from $\lambda_\Phi<0$ is obtained only in the $N_\nu=1$ case,
 while $\lambda_\Phi<0$ gives no constraints in the $N_\nu=2$ and 3 cases.
Thus, the constraints for $N_\nu=2$ and 3 are the same as obtained by the LHC experiments:
 $2.24\ (2.59)\,{\rm TeV} \lesssim M_{Z'}$,
 where the lower bound corresponds to the ATLAS (CMS) result.
On the other hand,
 we can find that
 the $Z'$ boson mass for $N_\nu=1$ is tightly restricted:
 $2.24\ (2.59)\,{\rm TeV} \lesssim M_{Z'} \lesssim 3.7\,{\rm TeV}$,
 where the upper bound is obtained by the condition of $\lambda_\Phi>0$.

\section{Conclusion} \label{sec:conclusion}

We have studied the scale invariant local $U(1)_\chi$ model
 with vanishing scalar potential at the Planck scale,
 which is the so-called flatland scenario.
The $U(1)_\chi$ symmetry is broken by the CW mechanism,
 and it subsequently leads to EW symmetry breaking.
Using the conditions for the CW mechanism to successfully occur and
 realize $M_h \simeq 125$\,GeV and $v_H \simeq 246$\,GeV,
 the physical quantities are uniquely determined once one parameter is fixed.

To constrain the physical quantities,
 we have investigated the vacuum stability using the two-loop RGEs.
First, we have considered $\lambda_H>0$ at all energy scales,
 and found that
 it is impossible to realize $M_h \simeq 125$\,GeV while keeping $\lambda_H>0$,
 the same situation as in the SM.
In the following results,
 we have given up $\lambda_H>0$ at any energy scale.

Next, we have considered $\lambda_\Phi>0$ at all energy scales.
When the number of relevant Majorana Yukawa couplings of the right-handed neutrinos is one,
 i.e., $N_\nu=1$,
 the lower bound of the $U(1)_\chi$ gauge coupling $g_\chi$ has been obtained
 by considering the decoupling effects of the $Z'$ boson and the right-handed neutrinos.
In practice,
 the condition $\lambda_\Phi >0$ is reasonable to consider
 $\lambda_\Phi(\mu<M_{Z'},M_N) \simeq \lambda_\Phi(M_{Z'}) \simeq \lambda_\Phi(M_N)>0$
 because of $\beta_{\lambda_\Phi} (\mu<M_{Z'},M_N) \simeq 0$.
Then, we have found 
 the lower bound of $g_\chi$, shown as the left shaded region in Fig.~\ref{result}.
However, the condition $\lambda_\phi>0$ does not constrain in the $N_\nu=2$ and 3 cases.
For $N_\nu=2$,
 the running of $\lambda_\Phi$ is monotonically and slowly decreasing
 from the EW scale to the Planck scale,
 quite untypically.
Thus, the condition $\lambda_\phi>0$ gives no constraint in the $N_\nu=2$ case,
 since $\lambda_\Phi$ is always positive. 
For $N_\nu=3$,
 the running of $\lambda_\Phi$ is similar to that for $N_\nu=1$,
 but the gradient of the running is much gentler.
Then, the $Z'$ boson and the right-handed neutrinos are decoupled
 before $\lambda_\Phi$ becomes negative even for $g_\chi \sim 0.01$.
Therefore, the small $g_\chi$ regions are almost not constrained in the $N_\nu=3$ case.

In addition, we have discussed the positive definiteness of the scalar mass squared eigenvalues.
The large $g_\chi$ generates the large scalar mixing,
 and it would make the lighter mass squared eigenvalue be negative.
Thus, it gives the upper bound of $g_\chi$,
 which is shown as the right shaded region in Fig.~\ref{result}.
As a result,
 considering the vacuum stability
 and the positive definiteness of the scalar mass squared eigenvalues,
 we have found the allowed parameter regions for the physical quantities
 as in Table \ref{table}.

Finally, we have mentioned the experimental bounds on $M_{Z'}$.
To obtain the constraints on $M_{Z'}$,
 we have discussed the following experiments:
 the deviation of the $\rho$-parameter from unity,
 the $pp$ collision to $e^+ e^-$ or $\mu^+ \mu^-$ at the LHC,
 and $e^+ e^- \rightarrow f\bar{f}$ at the LEP-II.
As a result, we have obtained the constraints shown in Fig.~\ref{all_const},
 and found that
 the $Z'$ boson mass for $N_\nu=1$ is tightly restricted
 to $2.24\ (2.59)\,{\rm TeV} \lesssim M_{Z'} \lesssim 3.7\,{\rm TeV}$,
 where the lower bound corresponds to the ATLAS (CMS) result.

\subsection*{\centering Acknowledgment} \label{Acknowledgement}
We thank S. Iso and Y. Orikasa for helpful discussions on the flatland scenario.
We also thank T. Yamashita for useful discussions and valuable comments on the RG analyses.
This work is partially supported by Scientific Grants
 by the Ministry of Education, Culture, Sports, Science and Technology,
 Nos. 24540272 and 26247038.
The work of Y.Y. is supported
 by Research Fellowships of the Japan Society for the Promotion of Science for Young Scientists
 (Grant No. 26$\cdot$2428).

\section*{Appendix}
\appendix
\section*{{\boldmath $\beta$ functions} in the $U(1)_\chi$ extended SM} \label{app:RGE}
The RGE of coupling $x$ is given by $dx/d\ln \mu = \beta_x$, in which $\mu$ is a renormalization scale.
The $\beta$ functions in the $U(1)_\chi$ extended SM are given by
\begin{eqnarray}
	\beta_{g_Y} &=& \frac{g_Y^3}{(4\pi)^2} \left[ \frac{41}{6} \right],\qquad
	\beta_{g_2} = \frac{g_2^3}{(4\pi)^2} \left[ -\frac{19}{6} \right],\qquad
	\beta_{g_3} = \frac{g_3^3}{(4\pi)^2} \left[ -7 \right],\\
	\beta_{g_\chi} &=& \frac{g_\chi}{(4\pi)^2} \left[ \frac{196}{25} g_\chi^2 + \frac{41}{6} g_{\rm mix}^2
		- \frac{4}{15} g_{\rm mix} g_\chi \right],\\
	\beta_{g_{\rm mix}} &=& \frac{1}{(4\pi)^2} \left[ g_{\rm mix} \left( \frac{41}{6} 
		\left( 2 g_Y^2 + g_{\rm mix}^2 \right) +\frac{196}{25} g_\chi^2 \right)
		- \frac{4}{15} g_\chi \left( g_Y^2 + g_{\rm mix}^2 \right) \right],\\
	\beta_{y_t} &=& \frac{y_t}{(4\pi)^2} \left[ \frac{9}{2}y_t^2
		- 8 g_3^2 - \frac{9}{4}g_2^2 - \frac{17}{12} \left( g_Y^2 + g_{\rm mix}^2 \right)
		- \frac{6}{25} g_\chi^2 + \frac{3}{5} g_{\rm mix} g_\chi \right], \label{yt}\\
	\beta_{y_{M_i}} &=& \frac{y_{M_i}}{(4\pi)^2} \left[ 4 y_{M_i}^2 + 2 {\rm Tr}(Y_M^2) - 6 g_\chi^2 \right],\\
	\beta_{\lambda_H}  &=& \frac{1}{(4\pi)^2} \left[ 
		\lambda_H \left( 24 \lambda_H + 12 y_t^2 - 3 \left( g_Y^2 + g_{\rm mix}^2 \right) 
		- 9 g_2^2 - \frac{48}{25} g_\chi^2 + \frac{24}{5} g_{\rm mix} g_\chi \right) \right. \nonumber\\
	&& \left.  + \lambda_{\rm mix}^2 - 6 y_t^4 + \frac{3}{8} \left( 2 g_2^4 + \left\{ g_2^2 + g_Y^2
		+ \left( g_{\rm mix} - \frac{4}{5} g_\chi \right)^2 \right\}^2 \right) \right],\\
	\beta_{\lambda_\Phi} &=& \frac{1}{(4\pi)^2} \left[
		\lambda_\Phi \left( 20 \lambda_\Phi + 8 {\rm Tr}(Y_M^2) -48 g_\chi^2 \right)
		+ 2 \lambda_{\rm mix}^2 -16 {\rm Tr}(Y_M^4) + 96 g_\chi^4 \right], \label{beta_lphi} \\
	\beta_{\lambda_{\rm mix}} &=& \frac{1}{(4\pi)^2} \left[ 
		\lambda_{\rm mix} \left( 12 \lambda_H + 8 \lambda_\Phi + 4 \lambda_{\rm mix} + 6 y_t^2 + 4 {\rm Tr}(Y_M^2)
		- 24 g_\chi^2 \frac{}{} \right. \frac{}{} \right. \nonumber \\
	&& \left. \left.
		- \frac{3}{2} \left\{ 3 g_2^2 + g_Y^2 + \left( g_{\rm mix} - \frac{4}{5} g_\chi \right)^2 \right\} \right)
		+ 12 \left( g_{\rm mix} - \frac{4}{5} g_\chi \right)^2 g_\chi^2 \right], \label{beta_lmix}
\end{eqnarray}
 up to the one-loop level.
We have only included the top quark Yukawa coupling,
 and omitted the other Yukawa couplings of the SM particles,
 since they do not contribute significantly to the Higgs quartic coupling and gauge couplings.
In this paper,
 we have used two-loop $\beta$ functions, which are obtained by SARAH \cite{Staub:2008uz}.

To solve the RGEs, we take the following boundary conditions \cite{Buttazzo:2013uya}:
\begin{eqnarray}
	&&g_Y(M_t) = 0.35761 + 0.00011 \left( \frac{M_t}{{\rm GeV}} - 173.10 \right),\\ 
	&&g_2(M_t) = 0.64822 + 0.00004 \left( \frac{M_t}{{\rm GeV}} - 173.10 \right),\\
	&&g_3(M_t) = 1.1666 - 0.00046 \left( \frac{M_t}{{\rm GeV}} - 173.10 \right) + 0.00314 \left( \frac{\alpha_3(M_Z) - 0.1184}{0.0007} \right),\\
	&&y_t(M_t) = 0.93558 + 0.00550 \left( \frac{M_t}{{\rm GeV}} - 173.10 \right) -0.00042 \left( \frac{\alpha_3(M_Z) - 0.1184}{0.0007} \right), \label{yt_mt}\\
	&&\alpha_3(M_Z) = 0.1184 \pm 0.0007,
\end{eqnarray}
 where $M_t$ is the pole mass of top quark.



\end{document}